\documentclass[
preprint,titlepage,showpacs,preprintnumbers,amsmath,amssymb,noshowkeys,superscriptaddress]{revtex4}
\usepackage{graphicx}
\usepackage{rotating}
\usepackage{epsfig}
\usepackage{color}
\usepackage{array}
\usepackage[dvipsnames]{xcolor}
\usepackage[utf8]{inputenc}
\usepackage[T2A]{fontenc}

\begin{document}
	
\title{Nuclear Fragmentation at Intermediate Energies in the DCM-QGSM-SMM Model}

\author{M.A.~Martemianov}
\email[E-mail: ]{mmartemi@gmail.com}
\affiliation{NRC «Kurchatov Institute», Moscow, Russia}%

\author{B.M.~Abramov}
\affiliation{NRC «Kurchatov Institute», Moscow, Russia}%

\author{S.A.~Bulychjov}
\affiliation{NRC «Kurchatov Institute», Moscow, Russia}%

\author{I.A.~Dukhovskoy}
\affiliation{NRC «Kurchatov Institute», Moscow, Russia}%

\author{V.V.~Kulikov}
\affiliation{NRC «Kurchatov Institute», Moscow, Russia}%

\author{A.A.~Kulikovskaya}
\affiliation{NRC «Kurchatov Institute», Moscow, Russia}
\affiliation{EANPO HE <<Moscow Technological Institute>>, Moscow, Russia}

\author{M.A.~Matsyuk}
\affiliation{NRC «Kurchatov Institute», Moscow, Russia}%

\begin{abstract}
The development of nucleus-nucleus interaction models is a rapidly developing area of heavy-ion physics. Recently, a new model DCM-QGSM-SMM, developed at JINR and oriented toward use within the NICA project at energies of several GeV/nucleon, became available. However, the mechanisms of nuclear interactions used in this model can potentially operate effectively at lower energies. In this paper, the model predictions are compared with the FRAGM and FIRST/GSI experimental data in the energy range of nucleus-nucleus interactions starting from 300 MeV/nucleon, as well as with the predictions of other models used in this energy range.
\end{abstract}


\maketitle

\section{INTRODUCTION}

Currently, there are a significant number of models of nucleus-nucleus interactions applicable in various energy ranges and for different interacting nuclei. Among the actively developing models, it is worth noting \mbox{DCM-QGSM-SMM} (Dubna Cascade Model -- Quark Gluon String Model -- Statistical Multifragmentation Model)~\cite{PaperDCMSMM1,PaperDCMSMM2}, which represents an approach that combines the advantages of the Dubna 
intranuclear cascade model (DCM) and the statistical multifragmentation model (SMM), allowing its use over a wide range of nuclear reactions and energies. This model is being developed at the Joint Institute for Nuclear Research and is focused on computer simulation and analysis of experimental data at the new BM@N~\cite{PaperBMN1} and MPD~\cite{PaperMPD} facilities of the NICA heavy-ion accelerator complex~\cite{PaperNICA} in the energy range 
of colliding nuclei above a few GeV/nucleon. 

The DCM-QGSM-SMM has no hard limitations on the energy range of applicability~\cite{PaperDCMSMM2}. 
However, the model was tested only in the region of comparatively high energies~\cite{PaperDCMSMM1, PaperDCMSMM2, PaperBMN2, PaperBMN3}. The aim of this work is to test the applicability of the DCM-QGSM-SMM model in the region of lower energies based on a comparison with the experimental data obtained at the FRAGM setup~\cite{PaperFRAGM1, PaperFRAGM2}. The model also will be compared with other widely used models in this energy range: 
the Binary Cascade (BC)~\cite{PaperBC} and the Liege Intranuclear Cascade (INCL)~\cite{PaperINCL}. Both models have been tested on experimental data in the intermediate energy range, with a particular focus on the carbon nuclei fragmentation. 

\section{MODEL DCM-QGSM-SMM}

The \mbox{DCM-QGSM-SMM} model is an extension of the \mbox{DCM-QGSM}, which is based on an upgraded version of the Dubna intranuclear cascade DCM~\cite{PaperDCM} and the quark-gluon string model QGSM~\cite{PaperQGSM}. The DCM model is a universal intranuclear cascade model for describing hadronic and nucleus-nucleus interactions. Cascade particles produced in primary binary interactions then pass through the nuclear matter medium and produce new secondary particles. The model also includes interactions of cascade particles with each other, taking into account experimental cross sections for elementary interactions. Upon the cascade process completion, single nucleons with similar coordinates and momenta produce light nuclei in accordance with the coalescence model. 

At energies above 5 GeV, binary collisions are considered within the framework of QGSM whose parameters are fixed using experimental data. The model considers the production and interaction of particles from the lightest SU(3) multiplets in the meson, baryon, and antibaryon sectors. In \mbox{DCM-QGSM}, after the end of the cascade part of the interaction of nuclei, when the produced particles have left the interaction region, residual nuclei and ejected nucleons are produced, which can combine into light fragments with atomic numbers up to four within the framework of coalescence model. 

The thermalization process of the residual nuclei occurs through a phase of pre-equilibrium emission of nucleons and light nuclei. Thermalized excited residual nuclei with atomic numbers up to 12 decay via the Fermi breakup mechanism, and heavier ones via the evaporation/fission mechanism. This approach to fragment production encountered serious difficulties in describing the interactions of heavy nuclei. Therefore, in the \mbox{DCM-QGSM-SMM} model, it was completely replaced by the statistical multifragmentation model. In this model, all excited residual nuclei formed at the intranuclear cascade stage are considered thermalized, which allows using the statistical multifragmentation model SMM~\cite{PaperSMM} for their decay. The resulting fragments decay further via the evaporation/fission mechanism. Actually, the main advantage of the DCM-QGSM-SMM is the wide range of masses and energies of colliding nuclei. 

\section{CARBON FRAGMENTATION DATA FROM THE FRAGM EXPERIMENT}

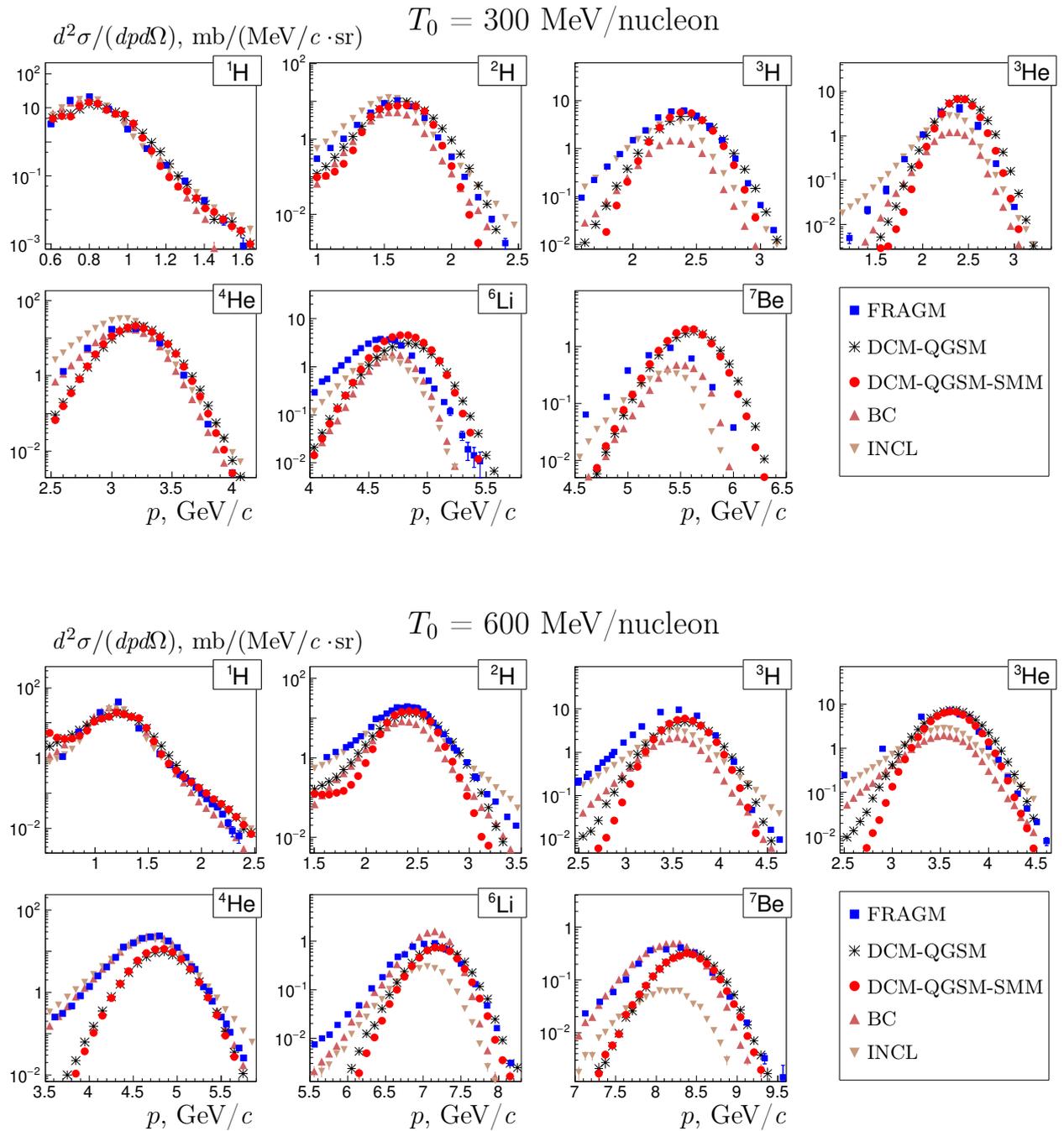
\begin{figure}[!htp]
\includegraphics[scale=0.83]{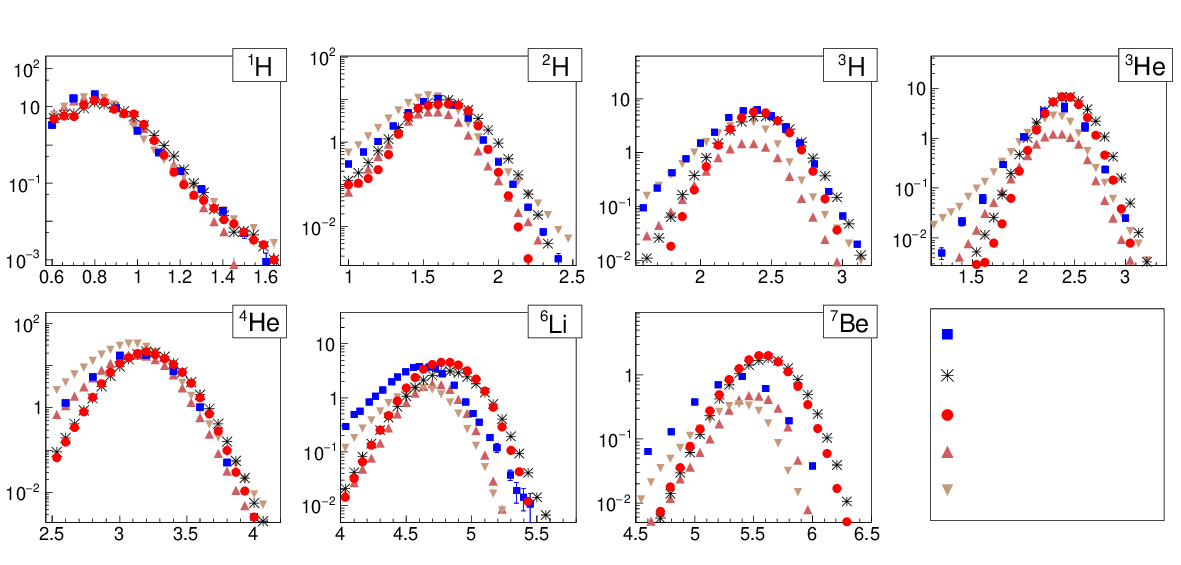}
\setlength{\unitlength}{5.0pt} 
\begin{picture}(100,0)
\put(36.,48.){\fontsize{14}{14}\selectfont{$T_0$ = 300 MeV/nucleon}}	
\put(4.2,46.7){\fontsize{11}{11}\selectfont{{\it d}$^{\hspace{0.1em}2}\sigma/$({\it dpd}$\Omega$), mb$/$(MeV$/${\it c}\hspace{0.1cm}$\cdot$\hspace{-0.1cm} sr)}}
\put(60.3,4){\fontsize{12}{12}\selectfont{$p$, GeV/{\it c}}}
\put(36.2,4){\fontsize{12}{12}\selectfont{$p$, GeV/{\it c}}}
\put(12.7,4){\fontsize{12}{12}\selectfont{$p$, GeV/{\it c}}}
\put(76.9,22.2){\fontsize{9}{9}\selectfont{FRAGM}}
\put(76.9,18.9){\fontsize{9}{9}\selectfont{DCM-QGSM}}
\put(76.9,15.7){\fontsize{9}{9}\selectfont{DCM-QGSM-SMM}}
\put(76.9,12.8){\fontsize{9}{9}\selectfont{BC}}
\put(76.9,9.8){\fontsize{9}{9}\selectfont{INCL}}
\end{picture}\vspace{0.8cm}
\includegraphics[scale=0.83]{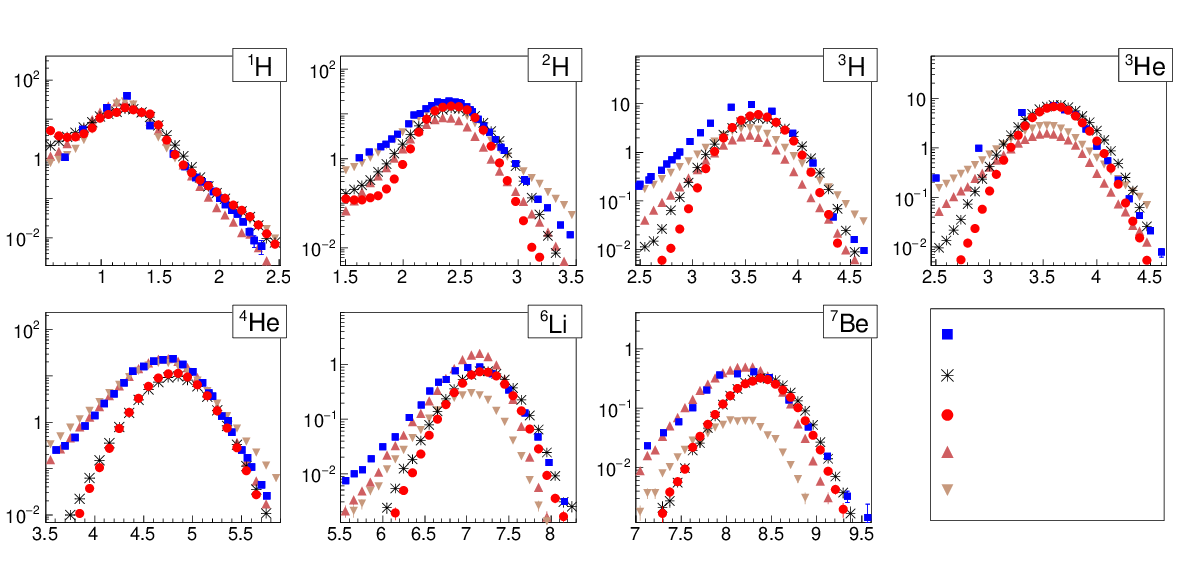}
\setlength{\unitlength}{5.0pt} 
\begin{picture}(100,0)
\put(36.,48.){\fontsize{14}{14}\selectfont{$T_0$ = 600 MeV/nucleon}}
\put(4.2,46.7){\fontsize{11}{11}\selectfont{{\it d}$^{\hspace{0.1em}2}\sigma/$({\it dpd}$\Omega$), mb$/$(MeV$/${\it c}\hspace{0.1cm}$\cdot$\hspace{-0.1cm} sr)}}
\put(60.3,4){\fontsize{12}{12}\selectfont{$p$, GeV/{\it c}}}
\put(36.2,4){\fontsize{12}{12}\selectfont{$p$, GeV/{\it c}}}
\put(12.7,4){\fontsize{12}{12}\selectfont{$p$, GeV/{\it c}}}
\put(76.9,22.2){\fontsize{9}{9}\selectfont{FRAGM}}
\put(76.9,18.9){\fontsize{9}{9}\selectfont{DCM-QGSM}}
\put(76.9,15.7){\fontsize{9}{9}\selectfont{DCM-QGSM-SMM}}
\put(76.9,12.8){\fontsize{9}{9}\selectfont{BC}}
\put(76.9,9.8){\fontsize{9}{9}\selectfont{INCL}}
\end{picture}\vspace{-0.8cm}
\caption{Differential cross sections $d^{2}\sigma/(dpd\Omega)$ of the light nuclei yield at $^{12}$C fragmentation 
with energies of 300 and 600 MeV/nucleon~\cite{PaperFRAGMData1, PaperFRAGMData2} 
in comparison with the predictions of four models of nucleus-nucleus interactions.}\label{Pict01}
\end{figure}	

\begin{figure}[!htp]
\includegraphics[scale=0.81]{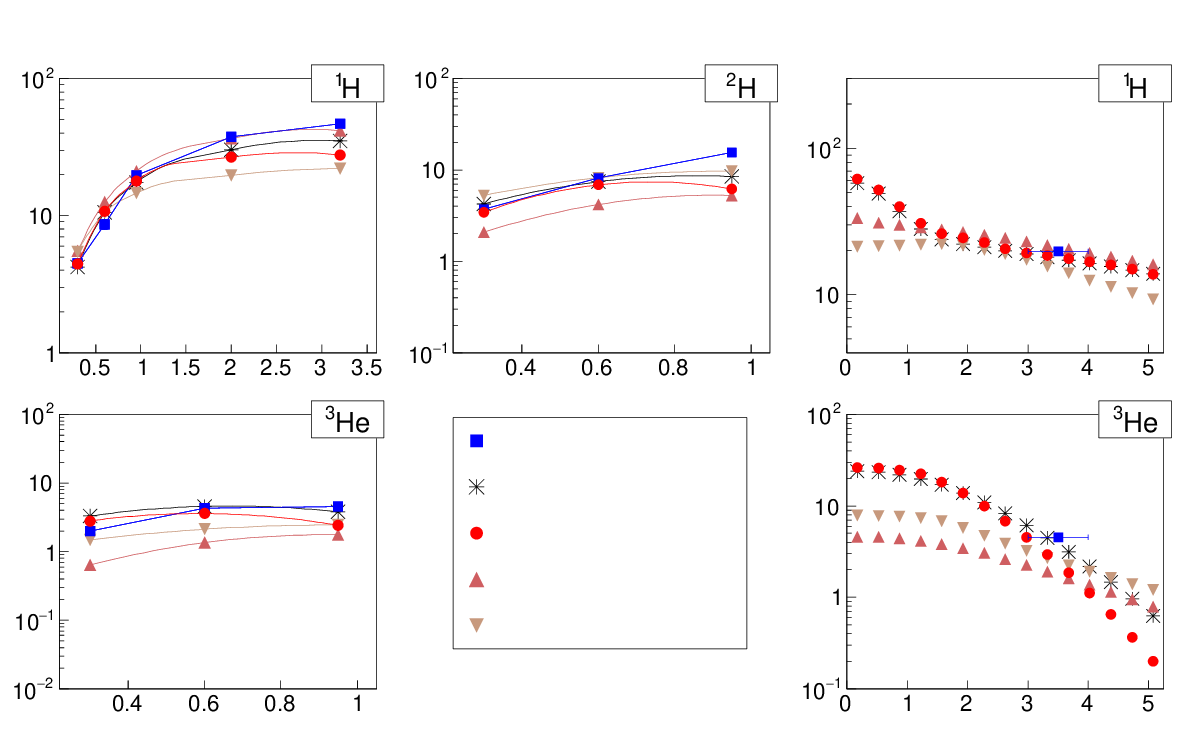}
\setlength{\unitlength}{5.0pt} 
\begin{picture}(100,0)
\put(6.0,58.1){\fontsize{11}{11}\selectfont{d$\sigma$/d$\Omega$, b/sr}}
\put(67.5,58.1){\fontsize{11}{11}\selectfont{d$\sigma$/d$\Omega$, b/sr}}
\put(37.,58.1){\fontsize{11}{11}\selectfont{d$\sigma$/d$\Omega$, b/sr}}
\put(84.6,5.5){\fontsize{12}{12}\selectfont{$\theta$, deg.}}
\put(47.6,31.6){\fontsize{11}{11}\selectfont{$T_0$, GeV/nucl.}}
\put(17,5.5){\fontsize{11}{11}\selectfont{$T_0$, GeV/nucl.}}
\put(70.6,60.6){\fontsize{11}{11}\selectfont{$T_0$ = 950 MeV/nucl.}}
\put(9,53.0){\fontsize{14}{14}\selectfont{{\it a}}}
\put(42,53.0){\fontsize{14}{14}\selectfont{{\it b}}}
\put(9,27.0){\fontsize{14}{14}\selectfont{{\it c}}}
\put(75,53.0){\fontsize{14}{14}\selectfont{{\it d}}}
\put(75,27.0){\fontsize{14}{14}\selectfont{{\it e}}}
\put(39.4,28.1){\fontsize{11}{11}\selectfont{FRAGM}}
\put(39.4,24.5){\fontsize{11}{11}\selectfont{DCM-QGSM}}
\put(39.4,20.9){\fontsize{11}{11}\selectfont{DCM-QGSM-SMM}}
\put(39.4,17.2){\fontsize{11}{11}\selectfont{BC}}
\put(39.4,13.8){\fontsize{11}{11}\selectfont{INCL}}
\end{picture}\vspace{-0.8cm}
\caption{Differential cross sections d$\sigma$/d$\Omega$ of the light nuclei yield at $^{12}$C fragmentation as function of the incident nucleus energy ({\it a}) - ({\it c}), as well as angular distributions for protons
({\it d}) and $^3$He~({\it e}) at $^{12}$C fragmentation with energy $T_{0}$~=~950~MeV/nucleon. 
The angular acceptance of the FRAGM setup is marked by a horizontal line.}\label{Pict02}
\end{figure}

Data on the fragmentation of carbon nuclei were obtained at the FRAGM experimental setup, created on the basis of heavy-ion accelerator-storage complex TWAC (Terra-Watt Accumulator)~\cite{PaperTWN}. Fragments formed 
at the interaction of carbon ion beam with an internal beryllium target were captured by a magneto-optical channel located at an angle of 3.5$^\circ$ to the accelerator's internal ion beam. Experimental data were collected at carbon nuclear energies from 300 to 3200~MeV/nucleon. Momentum spectra of the fragments were measured by scanning the rigidity of magneto-optical channel with a step of 50--100~MeV/{\it c}. Particle detection was performed by scintillation counters installed at both beamline foci. 

Fragment identification was accomplished by measuring time-of-flight and ionization losses in the scintillation detectors~\cite{PaperFRAGM3,PaperFRAGM4}. At low energies, the setup was capable of detecting all long-lived fragments, from protons to carbon isotopes. At energies above 1~GeV/nucleon, fragment detection was limited by the maximum rigidity of the magneto-optical channel of 6~GeV/{\it c}. 

Experimental data on the measured double-differential cross sections $d^2\sigma/(dpd\Omega)$ of the 
light nuclei yield (protons, $^2$H, $^3$H, $^3$He, $^4$He, $^6$Li, $^7$Be) at the fragmentation of carbon nuclei with energies of 300 and 600 MeV/nucleon as functions of the fragment momentum are shown in Fig.~\ref{Pict01}~\cite{PaperFRAGMData1,PaperFRAGMData2}. The plots show also the predictions of \mbox{DCM-QGSM}, \mbox{DCM-QGSM-SMM}, BC and INCL nucleus-nucleus interaction models. A similar comparison was partially carried out for 300 MeV/nucleon in~\cite{Paper_PPN}, but in Fig.~\ref{Pict01} fragments $^6$Li and $^7$Be were added and a comparison was also made with \mbox{DCM-QGSM} model. 

The proton momentum distributions are in good agreement with the predictions of all nucleus–nucleus interaction models. The \mbox{DCM-QGSM} and \mbox{DCM-QGSM-SMM} provide similar predictions across all models. They determine the fragment yield cross sections at the maximum of fragmentation peaks overall better than other models, but they also shift the center of this peak toward higher momenta and underestimate its width, which becomes more 
conspicuous with increasing fragment atomic number. Given the relatively wide spread of model predictions 
for different fragments, both among themselves and with experimental data, we can only conclude that they reflect the modern state of nuclear interaction modeling. 

Fig.~\ref{Pict02} shows the dependence of integrated yield for light fragments (protons, $^2$H, $^3$He) at 
an angle of 3.5$^{\circ}$ at fragmentation of $^{12}$C nuclei on a beryllium target for different energies of the incident nucleus: for protons in the range from 300 to 3200~MeV/nucleon, for $^2$H and $^3$He -- from 300 to 950~MeV/nucleon. Both experimental data and results of calculations in the \mbox{DCM-QGSM}, \mbox{DCM-QGSM-SMM}, BC and INCL models are presented. In general, good agreement with experiment can be clearly seen, especially in the low energy region. The \mbox{DCM-QGSM-SMM} model slightly underestimates the yields of $^2$H and $^3$He at an energy of 950 MeV/nucleon. This discrepancy may be due to the difference in angular dependencies of the fragment yields given in Fig.~\ref{Pict02} ({\it d, e}). It is evident, that the shapes of such dependencies differ for different models. The discrepancy between model predictions increases both with increasing incident nucleus energy and fragment mass. 

\section{CHARGED PIONS AT $^{12}$C FRAGMENTATION WITH AN ENERGY OF 3.2~GEV/NUCLEON}

Experimental data on the charged pions yields were obtained in the interaction of $^{12}$C with a beryllium target at an energy of 3.2 GeV/nucleon. To identify $\pi^{+}$ against the proton background, an additional Cherenkov counter placed in the second focus of the magneto-optical channel was used. In this case, the proton background suppression
was approximately 10$^3$, which, together with the time-of-flight measurement, made it possible to reliably select pions from the proton background and measure their momentum spectrum~\cite{PaperFRAGM5}. 

\begin{figure}[!htp]
\includegraphics[scale=0.87]{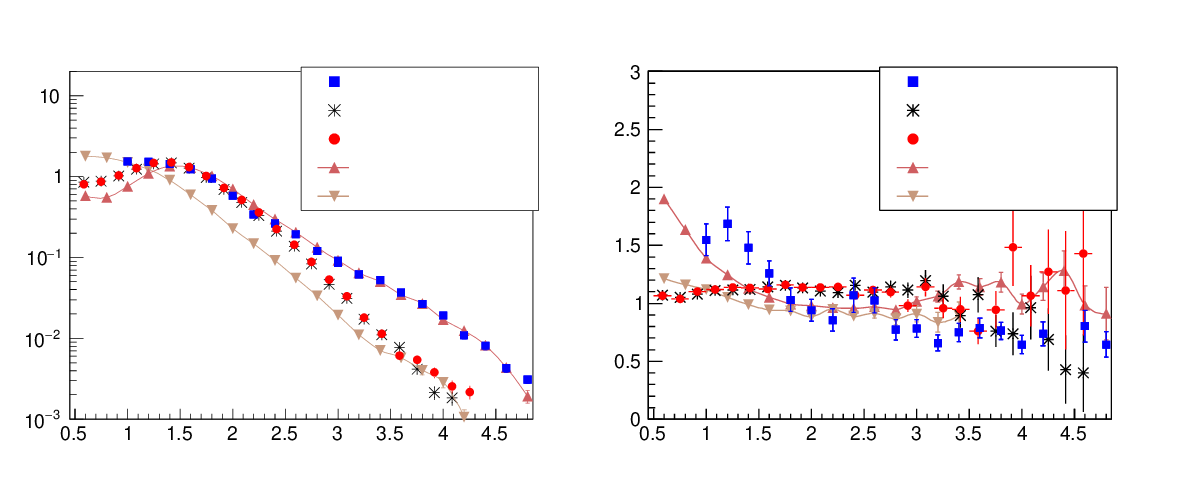}
\setlength{\unitlength}{5.0pt} 
\begin{picture}(100,0)
\put(6,40.){\fontsize{11}{11}\selectfont{{\it d}$^{\hspace{0.1em}2}\sigma/$({\it dpd}$\Omega$),
mb$/$(MeV$/${\it c}\hspace{0.1cm}$\cdot$\hspace{-0.1cm} sr)}}
\put(55,40.){\fontsize{12}{12}\selectfont{Ratio $\pi^{-}$~/~$\pi^{+}$}}
\put(14,34.0){\fontsize{14}{14}\selectfont{{\it a}}}
\put(64,34.0){\fontsize{14}{14}\selectfont{{\it b}}}
\put(78.5,37.1){\fontsize{8}{8}\selectfont{FRAGM}}
\put(78.5,34.6){\fontsize{8}{8}\selectfont{DCM-QGSM}}
\put(78.5,32.1){\fontsize{8}{8}\selectfont{DCM-QGSM-SMM}}
\put(78.5,29.8){\fontsize{8}{8}\selectfont{BC}}
\put(78.5,27.5){\fontsize{8}{8}\selectfont{INCL}}
\put(30.0,37.1){\fontsize{8}{8}\selectfont{FRAGM}}
\put(30.0,34.6){\fontsize{8}{8}\selectfont{DCM-QGSM}}
\put(30.0,32.1){\fontsize{8}{8}\selectfont{DCM-QGSM-SMM}}
\put(30.0,29.8){\fontsize{8}{8}\selectfont{BC}}
\put(30.0,27.5){\fontsize{8}{8}\selectfont{INCL}}
\put(33.6,4.8){\fontsize{12}{12}\selectfont\hspace{0.2cm}{{\it p}, GeV/{\it c}}}
\put(82.5,4.8){\fontsize{12}{12}\selectfont\hspace{0.2cm}{{\it p}, GeV/{\it c}}}
\end{picture}\vspace{-0.8cm}
\caption{Momentum distributions of $\pi^{-}$ ({\it a}) and the ratio of $\pi^{-}$ to $\pi^{+}$ ({\it b}) yields 
at $^{12}$C fragmentation with energy 3.2~GeV/nucleon. Experimental data are from papers~\cite{PaperFRAGM4,PaperFRAGM5}.
}\label{Pict03}
\end{figure}
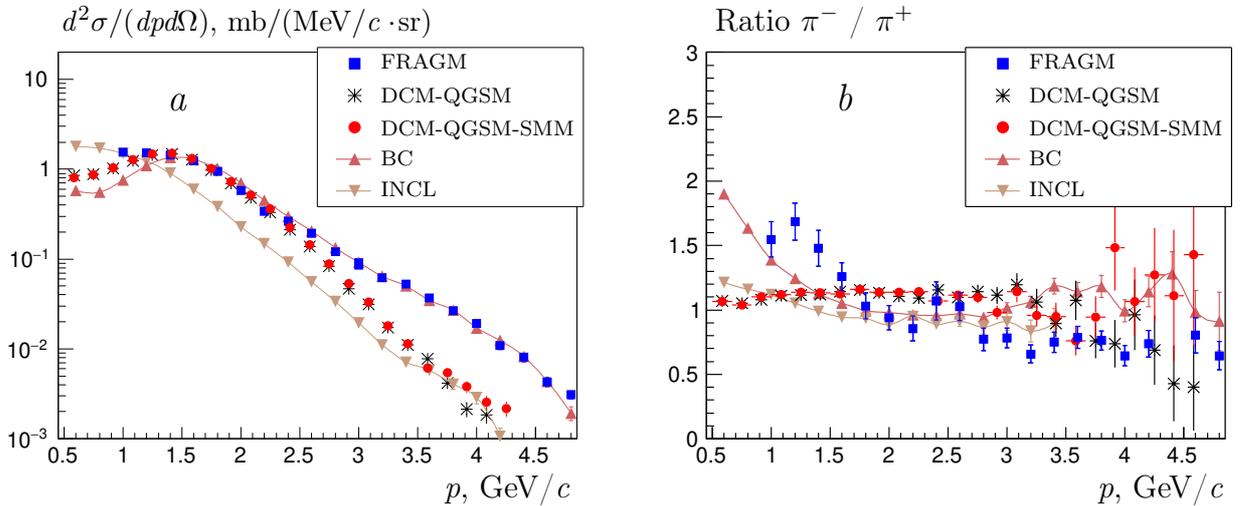

The differential $\pi^-$ yield cross section as a function of momentum is shown in Fig.~\ref{Pict03}~({\it a}) along with the predictions of four nucleus-nucleus interaction models. All models predict an exponential decrease in the cross section with increasing pion momentum, but with different slopes. The BC model gives the best agreement with the experimental data, and DCM-QGSM-SMM also satisfactorily describes the momentum spectrum up to 2.5 GeV/{\it c}. The ratio of the $\pi^-$ to $\pi^+$ yields as a function of their momenta is shown in Fig.~\ref{Pict03}~({\it b}) in comparison with the predictions of mentioned above models. 

The experimental data are in good agreement with the model predictions for pions with momenta above 1.5~GeV/{\it c}. Three models (\mbox{DCM-QGSM}, \mbox{DCM-QGSM-SMM} and INCL), determine the ratio value at the unity level for the entire energy spectrum under study. In the low momentum region, the BC model shows an increase in the $\pi^-$ to $\pi^+$ yield ratio with decreasing pion momentum, which is supported by experimental data, which may serve as an indication of the manifestation of the Coulomb effect in pion production at heavy-ion collisions. This effect was first observed in experiments on the Bevalac (USA) accelerator beams in pion production at small angles by $^{20}$Ne and $^{40}$Ar ions on different targets in the energy range from 400 to 500~MeV/nucleon and consisted of a sharp increase in the corresponding pion yield ratio at their velocities close to the velocity of incident nucleus~\cite{PaperCoulomb01, PaperCoulomb02}. Qualitatively, this is related to the action of the total charge of the spectators of incident nucleus, which attracts negatively charged pions, enriching the region of pion velocities close to the spectator velocity. 

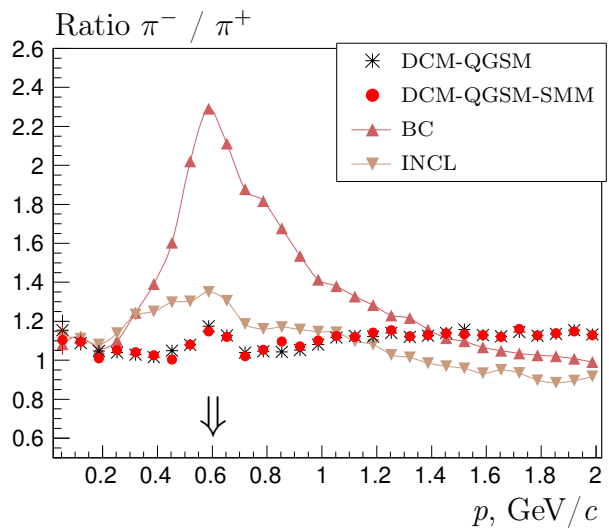
\begin{figure}[!htp]
\includegraphics[scale=0.46]{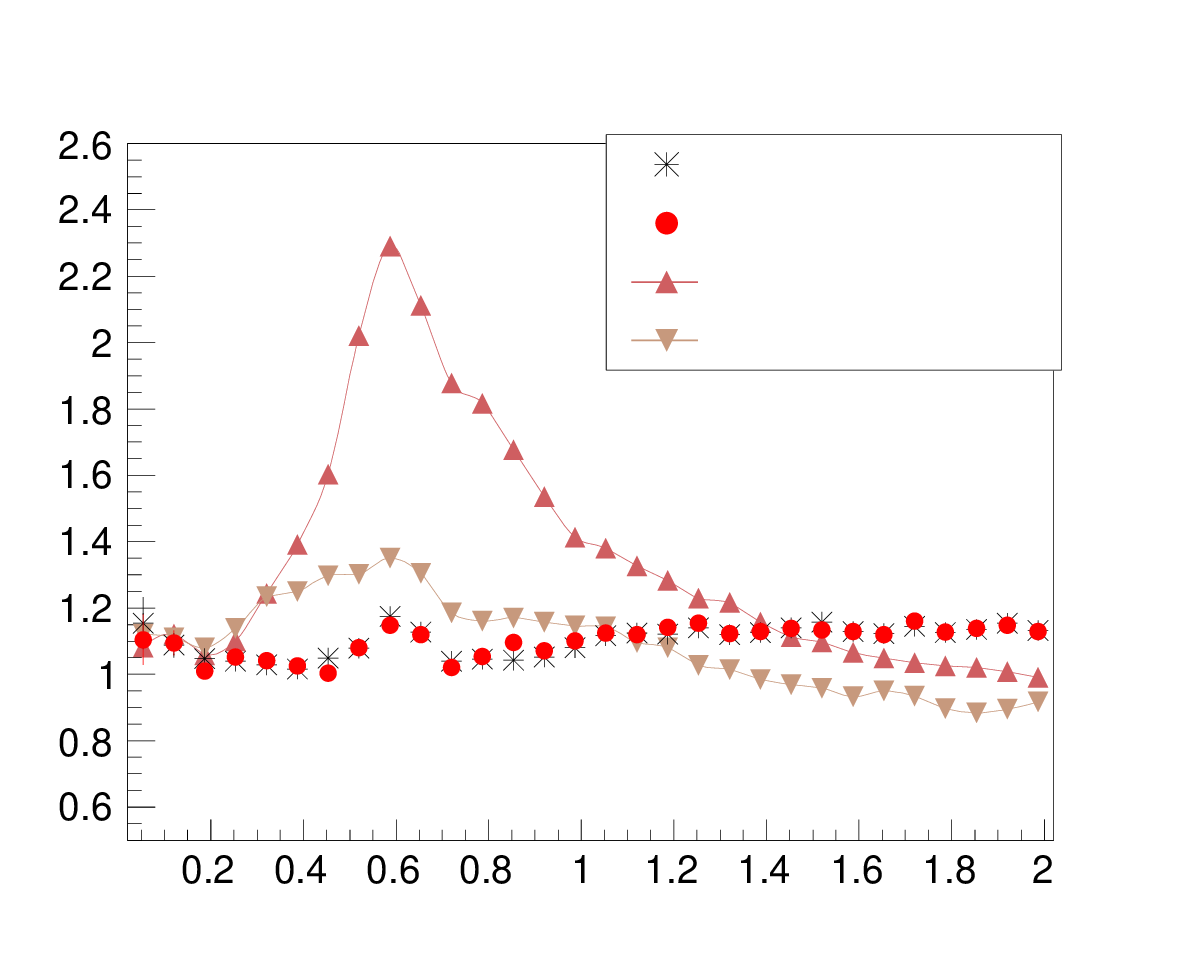}
\setlength{\unitlength}{3.0pt} 
\begin{picture}(100,0)
\put(16,69.){\fontsize{12}{12}\selectfont{Ratio $\pi^{-}$~/~$\pi^{+}$}}
\put(59.2,64.8){\fontsize{8}{8}\selectfont{DCM-QGSM}}
\put(59.2,60.5){\fontsize{8}{8}\selectfont{DCM-QGSM-SMM}}
\put(59.2,56.2){\fontsize{8}{8}\selectfont{BC}}
\put(59.2,52.0){\fontsize{8}{8}\selectfont{INCL}}
\put(68.0,8.0){\fontsize{12}{12}\selectfont{{\it p}, GeV/{\it c}}}
\put(34.0,19.5){\fontsize{16}{16}\selectfont{$\Downarrow$}}
\end{picture}\vspace{-0.8cm}
\caption{Momentum dependence of the ratio $\pi^{-}$ to $\pi^{+}$ yields at an angle less than $4^\circ$ 
at $^{12}$C fragmentation with energy 3.2~GeV/nucleon for various models of nucleus-nucleus interactions.
The arrow indicates a pion momentum corresponding to the velocity of incident carbon nucleus.}\label{Pict04}
\end{figure}		

Describing this effect within the framework of heavy-ion interactions encounters significant difficulties. A similar 
Coulomb effect was observed at high energies in the SPS (CERN) accelerator beams during the production of pions with small transverse momenta in Pb+Pb collisions at an energy of 158~A~GeV/{\it c}~\cite{PaperCoulomb03} and in Ar+Sc at 40~A~GeV/{\it c}~\cite{PaperCoulomb04}. Due to this, it would be interesting to see whether the discussed models of nucleus-nucleus interactions, oriented towards the intermediate-energy region, describe this effect. To this end, the model predictions were extended to lower pion momenta (see Fig.~\ref{Pict04}). The different model predictions, presented in the figure, show a structure with maximum in the region of 600 MeV/{\it c}, indicating that the Coulomb effect is taken into account in these models and that it manifests itself even for a small charge of the incident nucleus. It is evident that the predicted magnitudes of this effect vary considerably. Currently, the lack of experimental data does not allow testing these predictions. 

\section{ANGULAR DISTRIBUTIONS OF LIGHT FRAGMENTS}

To study the angular distributions of light fragments, carbon fragmentation data on a gold target from the FIRST experiment at the SIS heavy-ion accelerator complex at GSI~\cite{PaperFIRST, PaperExpFIRST} were used. Yields of light fragments were obtained for the carbon nuclei fragmentation with an energy of 400 MeV/nucleon. The experimental setup allowed the detection of fragments with energies from 100 to 1000 MeV/nucleon at various fragment emission angles. The differential cross sections d$\sigma$/$d\Omega$ were measured for a wide range of fragments, from protons to $^{11}$B, in the angular range from 0.2$^{\circ}$ to 5$^{\circ}$ at carbon fragmentation on a gold target. 

\begin{figure}[!htp]
\includegraphics[scale=0.81]{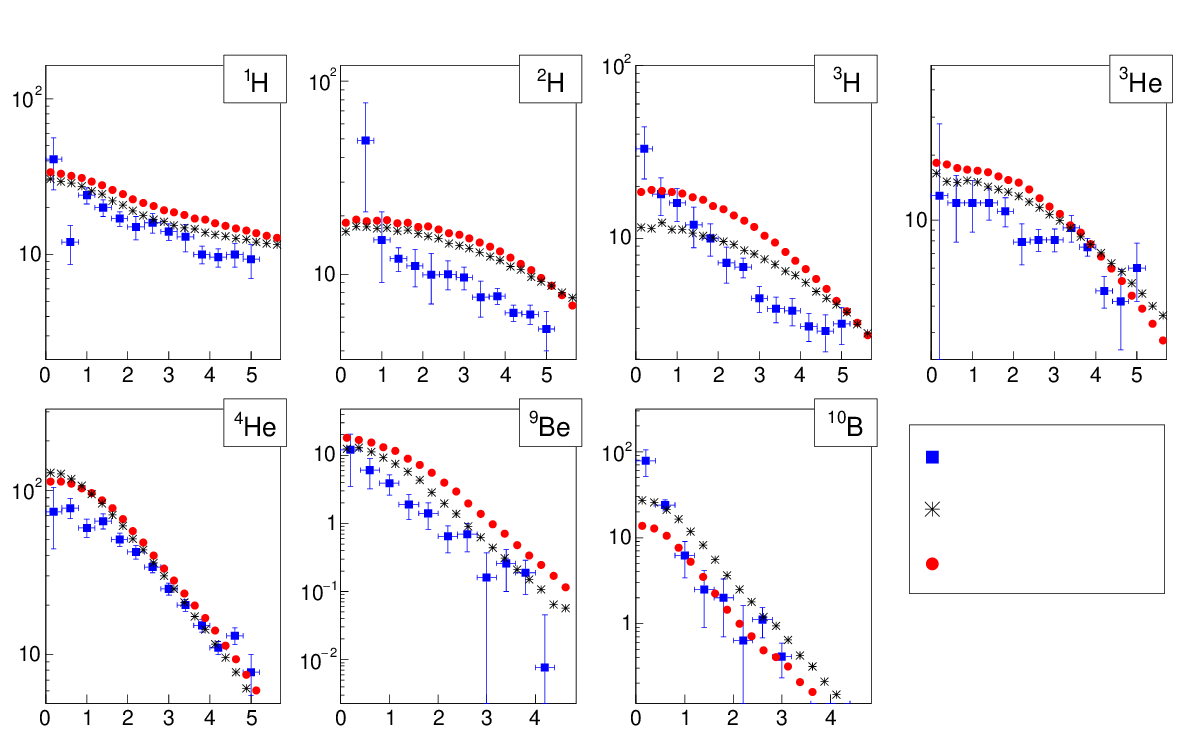}
\setlength{\unitlength}{5.0pt} 
\begin{picture}(100,0)
\put(4.5,59.4){\fontsize{11}{11}\selectfont{d$\sigma$/d$\Omega$, b/sr}}
\put(62,4.3){\fontsize{12}{12}\selectfont{$\theta$, deg.}}
\put(85,31.2){\fontsize{12}{12}\selectfont{$\theta$, deg.}}
\put(75.0,26.9){\fontsize{9}{9}\selectfont{FIRST/GSI}}
\put(75.0,22.9){\fontsize{9}{9}\selectfont{DCM-QGSM}}
\put(75.0,18.6){\fontsize{9}{9}\selectfont{DCM-QGSM-SMM}}
\end{picture}\vspace{-0.8cm}
\caption{Differential cross sections $d\sigma/d\Omega$ of the light fragments yield as function of the emission angle
at $^{12}$C fragmentation on $^{197}$Au target with 400~MeV/nucleon obtained in the experiment FIRST/GSI~\cite{PaperExpFIRST} and in the models DCM-QGSM, DCM-QGSM-SMM}\label{Pict05}
\end{figure}
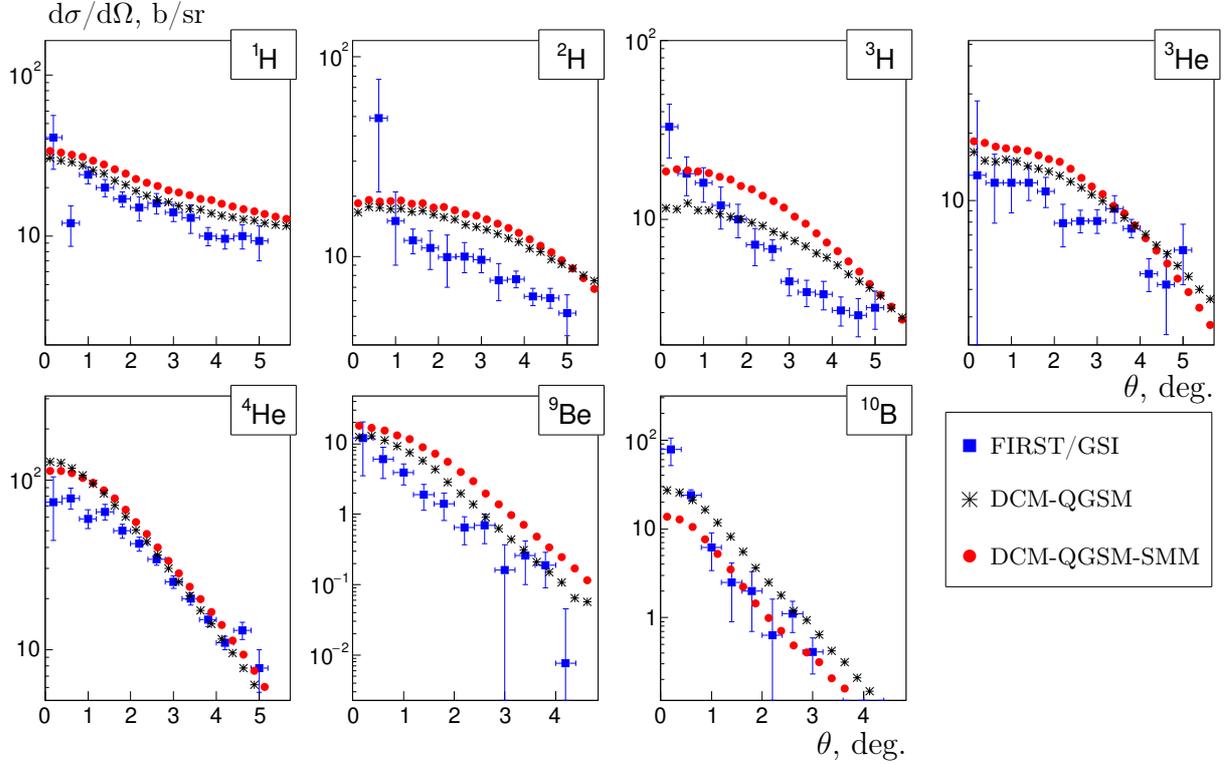

The experimental data for protons, $^2$H, $^3$H, $^3$He, $^4$He, $^9$Be, $^{10}$B are presented in Fig.~\ref{Pict05} together with the corresponding predictions of the \mbox{DCM-QGSM} and \mbox{DCM-QGSM-SMM} models. A significant advantage of these models is the absence of limitation on the mass numbers of colliding nuclei, 
which allows for the correct calculation of reactions with heavy targets such as gold. In general, satisfactory agreement between the model and experimental data can be observed. 

\section{CONCLUSION}

In general, it can be concluded that the \mbox{DCM-QGSM} model and its extended version \mbox{DCM-QGSM-SMM} correctly describe the fragmentation processes of carbon nuclei at intermediate energies. Comparison with the experimental data of FRAGM and FIRST/GSI demonstrated that both models reproduce the momentum and angular distributions of light fragments and charged pions at carbon fragmentation in the energy range from 300 to 3200~MeV/nucleon. 

This comparison demonstrates the applicability of \mbox{DCM-QGSM} and \mbox{DCM-QGSM-SMM} models at lower energies until 300~MeV/nucleon, where their prediction accuracy is comparable to that of models focused solely on intermediate energies. The observed minor difference between the \mbox{DCM-QGSM} and \mbox{DCM-QGSM-SMM} calculations for light fragments is expected, since the Fermi breakup mechanism and the SMM model give similar results for light nuclei. Extending this testing to reactions with heavier fragments and nuclei appears to be an important step in further
evaluation of the generality of these approaches. 

\section*{Acknowledgments}
The authors are grateful to G. Musulmanbekov for providing the computer code. We are also grateful to the TWAC accelerator complex staff and the technical staff of the FRAGM experiment for their significant contributions to the 
measurements.

\end{document}